\documentclass[
 reprint,
 amsmath,amssymb,
 aps,
]{revtex4-2}

\usepackage{graphicx}%
\usepackage{algorithmicx}%
\usepackage{algpseudocode}%
\usepackage{listings}%
\usepackage{hyperref}
\usepackage{ulem}
\usepackage{chngcntr}
\usepackage{float}
\usepackage{lineno}
\usepackage{xr-hyper} 

\begin{document}
\preprint{APS/123-QED}

\title[Article Title]{Resolved Anderson localization of light in dense three-dimensional dielectric disorder}

\author{Yevgen Grynko}
 \affiliation{BASF Coatings GmbH, Glasuritstraße 1, 48165 Münster, Germany}
 \email{yevgen.grynko@gmail.com}
 
\author{Jens Förstner} 
\affiliation{Department  of Theoretical Electrical Engineering, Paderborn University, Warburger Str. 100, Paderborn, 33098, Germany}
\email{jens.foerstner@uni-paderborn.de}

\date{\today}

\begin{abstract}
Strong localization of light in 3D dielectric disorder is challenging to establish because localized contributions can be hidden by diffusive leakage and finite-sample loss. We perform full-wave time-domain Maxwell simulations of dense high-index dielectric slabs and analyze late times after the diffusive component escapes. Transmission develops a non-exponential tail, and an effective diffusion coefficient approaches localized-dynamics scaling. Late-time spectra show narrow resonances with sub-unity mean Thouless conductance and Poisson-like spacings. Time- and frequency-resolved near fields reveal compact non-propagating modal domains separated by persistent low-intensity channels and interacting weakly. These converging signatures constitute finite-system evidence for resolved 3D Anderson localization, with the late-time field forming a self-organized, interference-defined confinement pattern.

\end{abstract}

\maketitle

Strong localization of electromagnetic waves in three-dimensional disordered media remains one of the most elusive problems in mesoscopic physics. First introduced by Anderson for electrons in random lattices \citep{Anderson1958}, this phenomenon describes the breakdown of diffusion due to coherent multiple scattering and interference. Later it was demonstrated for other kinds of waves,  in acoustics \cite{Hefei2008}, quantum systems \cite{Kondov2011, Crespi2013}, as well as 2D photonic \cite{Schwartz2007, Shubitidze24, Riboli2011, Mondal2019, Caselli2017} and microwave \cite{Laurent2007, Aubry2020, Dalichaouch1991} structures. For light, this problem is especially challenging in 3D, where vector effects, near-field coupling, polarization mixing, and finite-size leakage complicate unambiguous identification.
 Therefore, direct evidence for the strong localization of light in fully random 3D dielectric media remains controversial at present.  Experimental claims of localization in strongly scattering powders \cite{Wiersma1997,Stoerzer2006,Sperling2013} remain debated and have to be re-interpreted in the presence of absorption and inelastic  processes \citep{Scheffold1999, Scheffold2013, Sperling_2016, Skipetrov_2016}. 
 
Recent progress in high-performance computing and numerical time-domain methods for solving the Maxwell equations has revived the field by enabling large-scale 3D simulations deep into the late-time regime. In particular, first-principles simulations have shown clear 3D localization and an Anderson transition for vector electromagnetic waves in random ensembles of overlapping conducting spheres \citep{Yamilov_etal2022}. The same simulation framework reports an absence of localization for dielectric spheres even at very large index contrast, leaving the question of the 3D dielectric case open.  At the same time, numerous studies have reported spatially confined optical states and transport anomalies consistent with localization in correlated 3D \cite{Haberko2020}  and disordered 2D dielectric structures \cite{Schwartz2007, Shubitidze24, Riboli2011, Conley2014, Mondal2019}. These developments suggest that the absence of clear 3D evidence in dielectric systems may reflect not the impossibility of localization itself, but the difficulty of reaching and identifying the relevant regime.

In parallel, localization landscape (LL) theory has emerged as a framework for describing wave confinement in complex media \cite{FilocheMayboroda2012}.   Extended from scalar continuous systems to tight-binding and 3D cases  \cite{Lyra2015,Skipetrov2024,Filoche2024}, it predicts valleys that partition the medium into localization basins. Its optical relevance has been shown in 2D scalar systems \cite{Shubitidze24}, and connected LL barriers have been proposed as a spatial indicator of the mobility edge \cite{Filoche2024}.


Beyond its fundamental importance, 3D localization in dielectric media could extend disorder-based photonic control to fully three-dimensional systems, with possible implications for random lasing, cavity QED, sensing, optomechanics, reservoir computing, and nonlinear photonic processing \cite{Rashidi21,Vasco_Hughes2018,Trojak2017,Arregui2023,Rafayelyan2020,Wang2024}.



In this work, we use large-scale time-domain simulations to study light propagation through dense disordered slabs of high-index dielectric particles. Going beyond transport-based diagnostics, we directly resolve the real-space structure of localized dynamics within a 3D vector Maxwell system. Our results uncover a late-time regime in which long-lived confined electromagnetic modes emerge together with established transport and spectral signatures associated with Anderson localization. 


\begin{figure}[htbp]
  \centering
  \includegraphics[width=4cm]{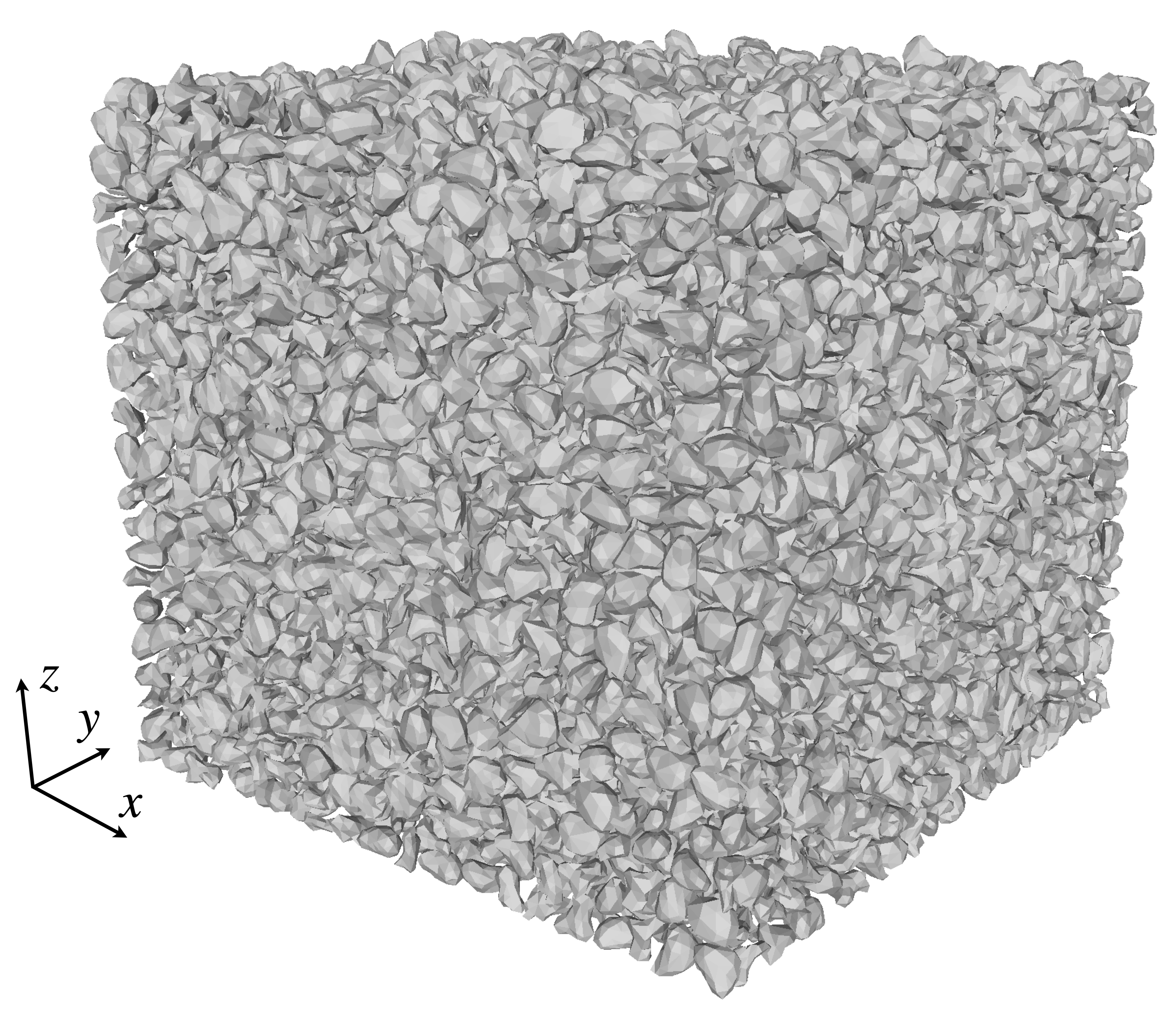}
  \caption{Example of a unit cell with thickness $X_L=16$ filled with 9000 irregular particles at volume fraction  $\rho = 0.44$. The particles have refractive index $n=3$ and characteristic size parameter $kr \approx 1$. The transverse directions XY are periodic, while the slab is open along the propagation direction Z.}
  \label{3D_sample}
\end{figure}

We apply the discontinuous Galerkin time-domain method (DGTD) \cite{HESTHAVEN2002} to solve the full-wave electromagnetic problem for systems of densely packed dielectric particles. 
A previous finite-system study \citep{grynko2023} showed that for particles with refractive index $n=3$ and size parameter $kr \sim 1$ ($k$ is the wavenumber and $r$ the radius of the circumscribing sphere), progressive dense packing leads to a transition in transport observables at volume fractions $\rho > 0.4$, consistent with an approach to localization. This behavior was observed for granular samples with different sizes and boundary conditions, whereas lower-index or sparser systems remained diffusive. Here we use the same numerical setup and parameters and simulate light transmission by periodic monodisperse slabs with 6000--12000 non-overlapping particles with preserved granularity, mimicking a realistic powder (Fig. \ref{3D_sample}). Gaussian random field shapes are used as constituents \citep{Grynko2020, Grynko_chapter_2022, Grynko2018}. The volume fraction is fixed at $\rho = 0.44$ and refractive index at $n=3$. For the main study, we consider a unit cell with dimensions $X_d = 18$ (in size parameter units) along the X and Y axes and thickness $X_L = 16$  along Z, with $N=9000$  particles and four disorder realizations. Additionally, we simulated transmission by samples with $X_L=10.6$ and $X_L=21.3$  to study the effect of layer thickness (Supplemental Material, Sec. \ref{Thickness_test}). In all simulations, a model sample is illuminated by a short broadband $E_x$-polarized pulse propagating along the Z axis. More details on the numerical model are provided in the Supplemental Material, Sec. \ref{Numerical_method}.

 \begin{figure}[htbp]
  \centering
  \includegraphics[width=8.5cm]{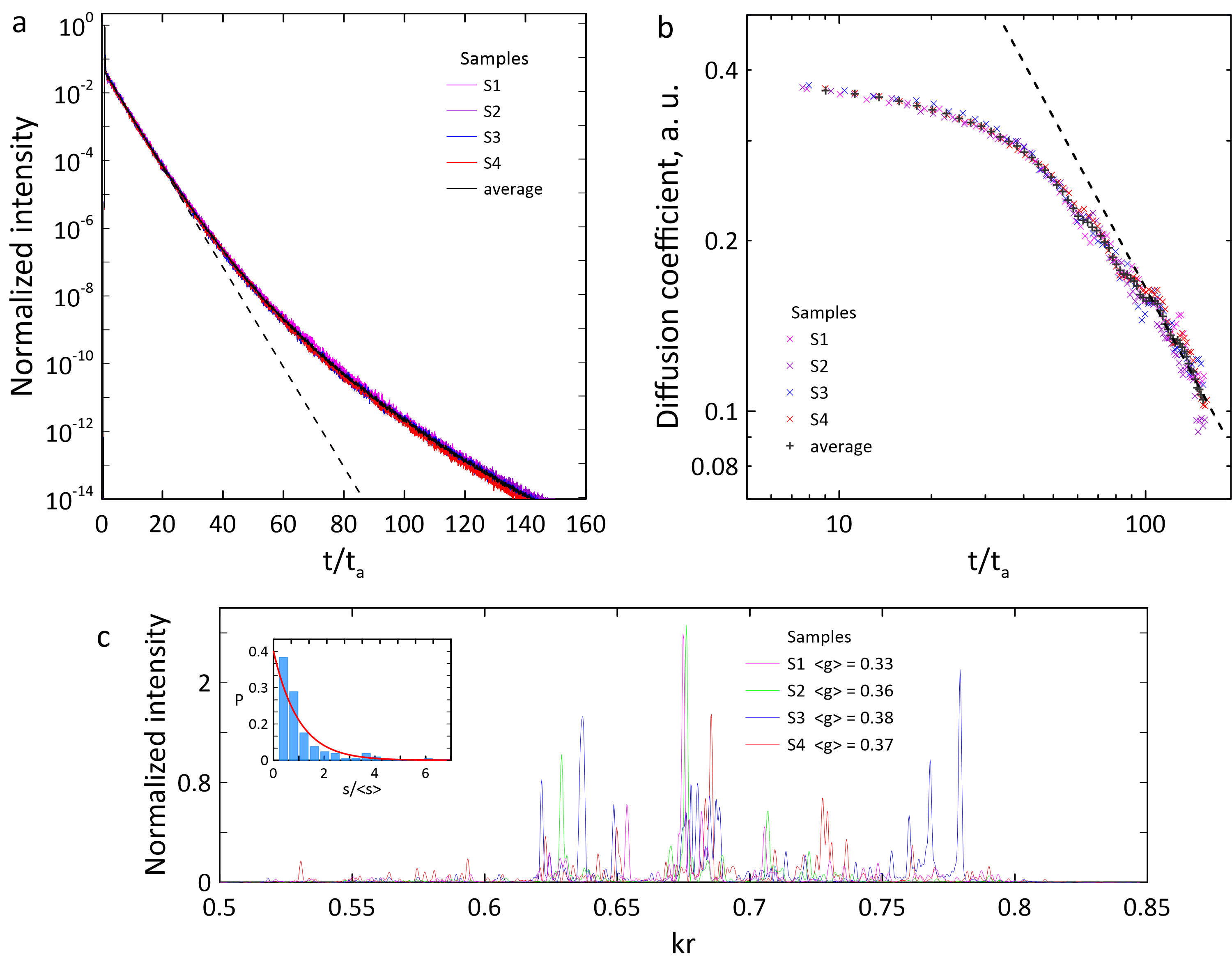}
  \caption{a, Time-resolved transmission $T(t)$ for four independent disorder realizations with slab thickness $X_L=16$. The dashed line shows an exponential fit to the early-time regime.
  b, Effective time-dependent diffusion coefficient $D(t)$ extracted by local exponential fitting of $T(t)$. The average $D(t)$ is obtained by fitting the $T(t)$ curve averaged over four samples. The dashed line shows the $1/t$ dependence expected for localized dynamics in an open system.
  c, Late-time ($t>130t_a$) transmission spectra and Thouless conductances $\langle g\rangle$ measured for four samples. The horizontal axis is given in terms of the particle size parameter $kr$; thus, the central wavelength of the excitation pulse corresponds to $kr=1$. The inset shows the normalized spacing distribution obtained by pooling the resonance spacings from all four spectra; its approximately Poissonian form is consistent with weakly overlapping localized modes.}
  \label{Tt_Dt_Spectra}
\end{figure}

We first analyze the time-resolved transmission $T(t)$ and the corresponding effective time-dependent diffusion coefficient $D(t)$ for four disordered samples. At sufficiently long observation times, we observe a qualitative transition in the transmission characteristics. Fig. \ref{Tt_Dt_Spectra}a shows the emergence of a non-exponential tail in $T(t)$ after a diffusive period at early times characterized by an exponential dependence. The time scale is normalized by the pulse peak arrival time $t_a$. As a consistency check, the transport parameter $kl^*_{eff}$ ($l^*_{eff}$ is the effective mean transport length for a given unit cell) for the diffusive component of the propagating field can be estimated from the slope of ln$T(t)$ in the early-time exponential regime \cite{LAGENDIJK1996}. If $s$ is the absolute value of the slope, with known slab thickness $L$,  we can calculate the diffusion coefficient $D \approx L^2 s / \pi^2$.  Using $D$ and the effective refractive index $n_{eff}\approx 1.87$  the transport parameter becomes $kl^*_{eff} \approx 3 k n_{eff} D / c \approx 1.21$. This is consistent with the condition $kl^* \sim 1 $ for the strong scattering regime \cite{IoffeRegel1960, LAGENDIJK1996, SkipetrovSokolov2018}. However, we note that such an estimation is based on the averaged spectral content of the broadband pulse and, therefore, $kl^*_{eff}$ is an effective broadband transport parameter here. If applied to localization, this condition should be considered as qualitative, which follows from the theoretical analysis \cite{SkipetrovSokolov2018} and experimental results, e.g., \cite{Mondal2019}.

The effective diffusion coefficient $D(t)$ obtained by local exponential fitting of the $T(t)$ curves further confirms the transition (Fig. \ref{Tt_Dt_Spectra}b). After $t \approx 40 t_a $, $D(t)$ progressively becomes time- and sample-dependent. We note that the average $D(t)$ curve is deduced from the average $T(t)$ signal. The pronounced step-like variations of $D(t)$ observed for individual disorder realizations are strongly reduced after averaging and, at $t \gtrsim 100t_a$, the average $D(t)$ reaches the $1/t$ dependence in accordance with the analytical prediction for the localized regime \cite{SkipetrovVanTiggelen2006}. 

Early-time $D(t)$ corresponds to the diffusive process, where self-averaging makes the integral transmission $T(t)$ nearly sample independent. At late times, after faster-decaying extended components have escaped, the signal is dominated by a limited number of long-lived, spatially confined quasi-modes with different leakage rates. Such modes are expected to produce an extended $D(t) \sim 1/t$ regime in the localized phase \cite{SkipetrovVanTiggelen2006, Yamilov_etal2022}. In a finite sample, however, the number of contributing modes is small, so individual modal handovers can appear as step-like variations of $D(t)$. When one bright mode dominates the transmitted signal, the apparent decay rate may temporarily flatten or deviate from the asymptotic trend; after this contribution fades, the localized scaling is recovered. Averaging $T(t)$ over realizations mixes different modal lifetimes and coupling strengths, washing out these realization-specific crossovers and producing a smoother $D(t)$.

The late-time transmission spectra, evaluated after the decay of the initial transient regime  ($t > 130t_a$), are in full agreement with this interpretation. They are dominated by narrow resonances with strongly fluctuating amplitudes but comparatively similar linewidths (Fig. \ref{Tt_Dt_Spectra}c). Remarkably, for four independent disorder realizations of the same sample thickness, the long-lived resonances are concentrated in nearly the same narrow spectral interval, $kr \simeq 0.62 - 0.78$, indicating a late-time spectral window in which leakage is reduced and spatial confinement is efficient.

This behavior is consistent with modal localization in an open system. The linewidth of a localized resonance is mainly set by boundary leakage, which depends on confinement length, position within the slab, and hybridization with nearby modes. Although these factors vary between modes, they remain sufficiently similar within the late-time localized band to produce a relatively narrow distribution of decay rates. By contrast, resonance amplitudes depend on the overlap of each mode with the incident field and detection monitor. As localized modes occupy different regions in the bulk, this overlap fluctuates strongly, so modes with comparable lifetimes can have very different spectral weights.

The spectral isolation of the modes can be quantified using the Thouless conductance $g$ \cite{THOULESS197493}, defined as the ratio of resonance linewidths to nearest-neighbor spacings, together with the level-spacing statistics. Here, we use a mean value $\langle g\rangle$ as a spectral-overlap parameter for the finite open system. For four independent realizations, the extracted values remain close,  $\langle g\rangle = 0.33–0.38$, demonstrating that the system is in the weak-overlap regime $\langle g\rangle < 1$. The normalized spacing distribution $P(s/\langle s\rangle)$ constructed from 142 nearest-neighbor spacings pooled over the four spectra is shown in the inset of Fig. \ref{Tt_Dt_Spectra}c. Its approximately Poissonian form is consistent with weakly interacting localized modes \cite{EscalanteSkipetrov2018}, in contrast to the level repulsion and Wigner-Dyson statistics expected for overlapping extended modes. Thus, the late-time spectral band, its reproducibility across independent samples, the sub-unity Thouless conductance, and the Poisson-like spacing statistics provide mutually consistent spectral evidence for a multimode localized regime.

\begin{figure}[htbp]
  \centering
  \includegraphics[width=8cm]{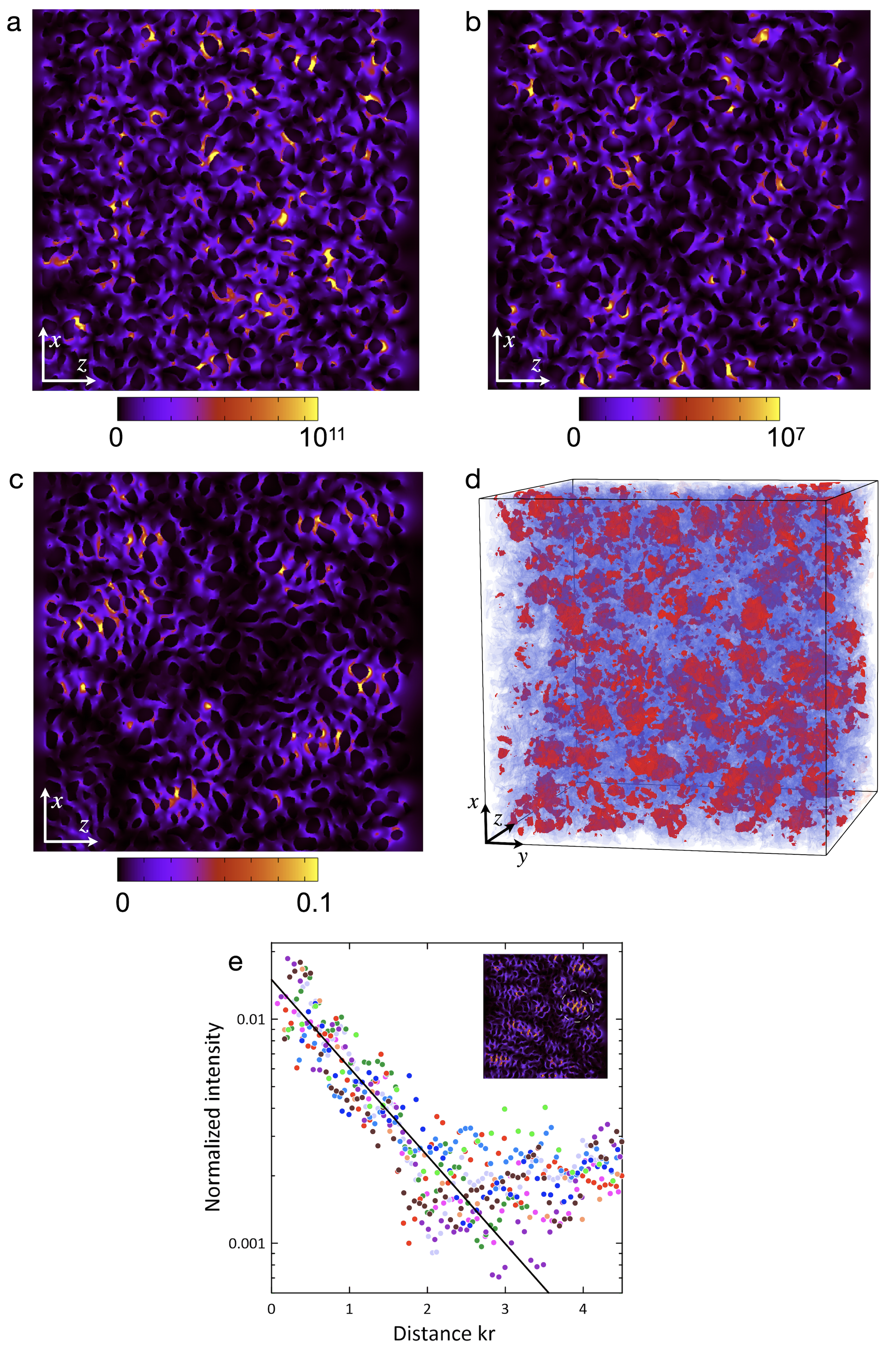}
  \caption{Real-space emergence of spatially isolated late-time modal domains. a--c, Instantaneous near-field intensity ($|E|^2$) maps in the XZ plane at representative times. At early times (a, $t=10t_a$), the field consists of randomly distributed hotspots characteristic of a diffusive regime. At intermediate times (b, $t=40t_a$), structured regions begin to emerge as the fastest extended components decay. At late times (c, $t=140t_a$), the field breaks into compact, non-propagating intensity clusters. d, 3D rendering of the late-time near field. High-intensity thresholding reveals isolated modal cores (red), while a lower-intensity isosurface (blue) shows the surrounding weak field network. e, Average radial intensity profiles for ten compact, comparatively isolated late-time modes. The inset shows an example of such a mode. The black line corresponds to an exponential fit.}
  \label{3_field_maps_and_3D}
\end{figure}

A transition in transport characteristics is expected to reflect the near field evolution leading to domination of long-lived modes. Fig. \ref{3_field_maps_and_3D} shows representative instantaneous XZ intensity maps recorded at different times. Fig. \ref{3_field_maps_and_3D}a corresponds to the early times when $T(t)$ is exponential ($t \approx 10t_a$) and shows a randomly scattered distribution of the field hotspots as expected for classic diffusion. Fig. \ref{3_field_maps_and_3D}b corresponds to a transient regime of small deviation of $T(t)$ from exponential decay and emerging structure in the near field ($t > 40t_a$). We previously observed similar behavior in smaller slabs with the same particle parameters and volume fraction \cite{grynko2023}: randomly scattered hotspots tend to form structured patterns with time, reaching the stage with pinned modes of increased lifetimes  but remaining spatially overlapped. Here, we extended observation in larger samples to much longer times and reach a regime in which persistent non-propagating modal domains become spatially isolated at $t > 100t_a$ (Fig. \ref{3_field_maps_and_3D}c). This provides a direct real-space indicator of strong light localization. Fig. \ref{3_field_maps_and_3D}d shows a 3D representation of such a late-time near field. Thresholding the high-intensity regions reveals randomly distributed isolated blobs of nearly the same spatial scale, corresponding to the bright cores of localized modes (red color). A lower intensity background network is shown with the blue 3D contour. Interestingly, the number of bright non-overlapping modal cores in such 3D snapshots is typically $\sim 30-50$, comparable to the number of transmission peaks in the spectra. Here, the strongest transmission resonances correspond to a small modal subset that combines long lifetimes with efficient coupling to the external transmission channel, while many other localized modes remain weakly visible in the spectra despite being clearly present in the internal field. This dynamical picture is illustrated in the late-time field-map Videos 1 and 2, while cycle-averaged maps and correlation analysis confirm that the near-field confinement pattern persists over hundreds of optical cycles (see Supplemental Material, Sec.~\ref{Supplement_video}).

 We interpret dark separation valleys as regions of coherent field cancellation produced by the superposition of localized modes and residual background contributions. This interpretation is supported by their weak connection to the material geometry: rather than following particle-air interfaces, as expected for single-particle resonances, the dark channels often cut through dielectric material. The mode boundaries are therefore defined by the coherent field structure rather than by individual scatterers. 

An analysis of ten comparatively isolated mode profiles selected from the instantaneous field maps shows that, despite noticeable mode-to-mode fluctuations, their short-range spatial decay follows a reproducible exponential trend (Fig. \ref{3_field_maps_and_3D}e). For this analysis, we chose compact modal cores whose nearest bright neighboring core was separated by $d_{\mathrm{nn}}\gtrsim 4-5$ in the size parameter units. The scatter increases at larger distances, where modal tails are more sensitive to disorder, residual modal overlap, and weak background contributions. 
Fitting the decay with $I(r)\propto \exp(-2r/\xi)$ gives $k\xi/2 \approx 2.21$, corresponding to $\xi \approx 0.7\lambda$ in free space and $1.31\lambda_{\mathrm{eff}}$ in the medium. Because the instantaneous field is a coherent superposition of several long-lived quasi-modes, this fitted length characterizes an effective modal-core confinement scale rather than an eigenmode localization length. This reproducible trend therefore provides an additional real-space indicator of a localized resonant regime. Wavelength-scale spatial decay of compact electromagnetic modes has also been reported in strongly scattering two-dimensional photonic and microwave systems \cite{Laurent2007, RazoLopez2024Localization, Caselli2017, Riboli2011, Molinari2012, Xianling2014}.

Thus, the qualitative transitions observed in the near field are fully consistent with the changes in the integral transport observables, \(T(t)\) and \(D(t)\), as well as with the late-time transmission spectra. Taken together, and including the reproducibility for different slab thicknesses (Supplemental Material, Sec. \ref{Thickness_test}), these signatures provide consistent evidence for Anderson-localized electromagnetic modes in realistic disordered dielectric media. A precise determination of the mobility edge would require a dedicated finite-size scaling analysis \cite{Abrahams1979}; nevertheless, the simultaneous onset of anomalous long-time transport, sub-unity Thouless conductance, and spatially confined near-field patterns is consistent with the late-time spectral window lying on its localized side.

\begin{figure}[htbp]
  \centering
  \includegraphics[width=8cm]{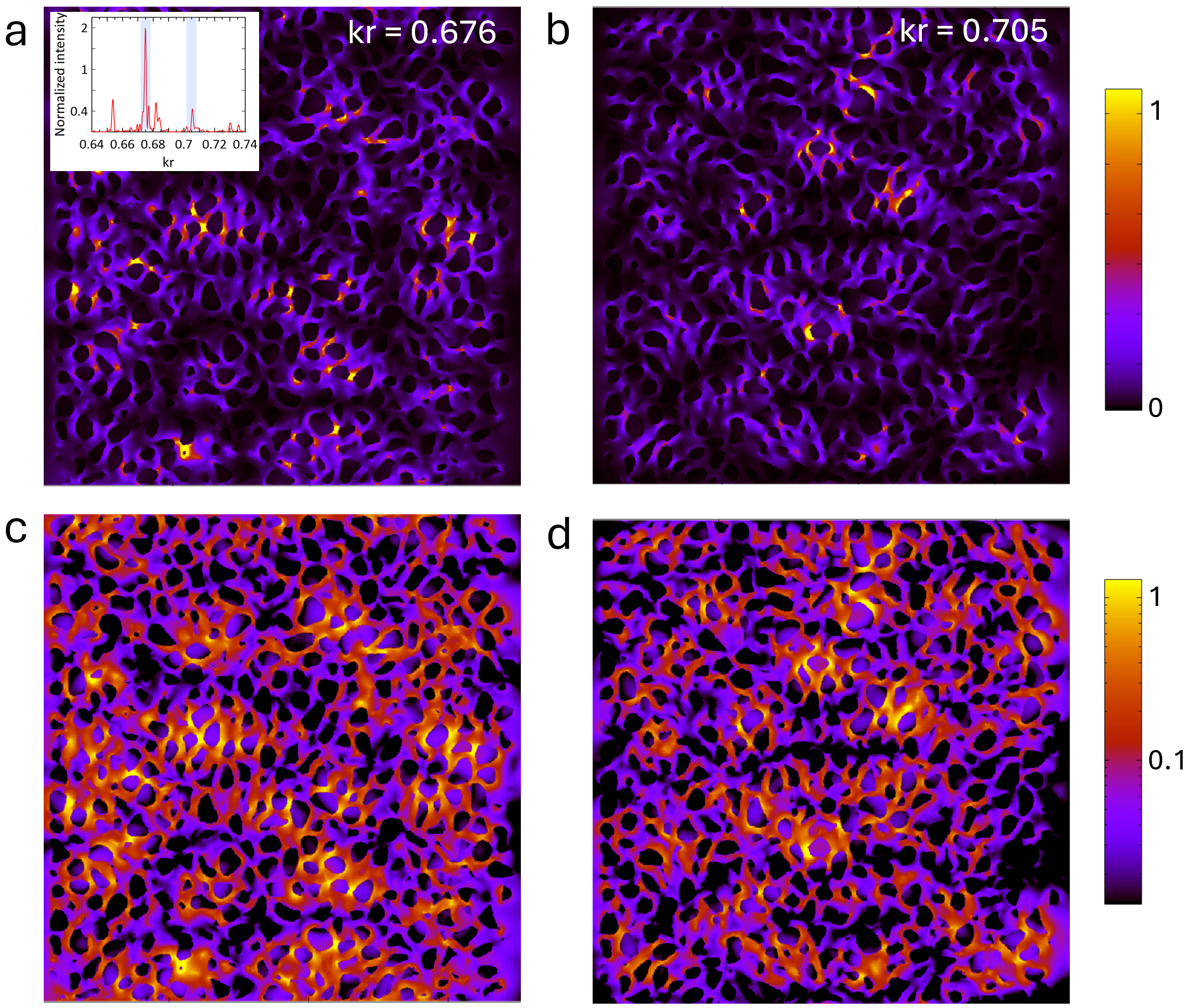}
  \caption{Late-time ($t > 130 t_a$) Fourier-transformed near-field YZ maps computed for sample S1 for two frequency windows shown in the inset spectrum. a,b, Near field maps corresponding to windows centered at $kr = 0.676$ and 0.705, respectively. c, d, The same maps shown on a logarithmic color scale.}
  \label{frequency_maps}
\end{figure}

Frequency-resolved near-field maps are consistent with this picture. Fig. \ref{frequency_maps} shows representative cross-sectional maps of the late-time Fourier-transformed near field for two spectral windows for sample S1, centered at kr=0.676 and kr=0.705, respectively. Because the Fourier transform is evaluated over a finite time interval, each map represents a narrow frequency window and may contain several nearby resonances. The shaded regions in Fig.~\ref{frequency_maps}a, inset, indicate the corresponding spectral width of $\Delta(kr) \approx 0.006$.

The windows select different subsets of long-lived modal components. For the window centered at kr=0.676 (Fig. \ref{frequency_maps}a), where several close resonances contribute, the near field shows a partially hybridized localized subset: several compact cores are weakly connected, but the field remains fragmented and does not form a system-spanning channel. By contrast, the window centered at kr=0.705 (Fig. \ref{frequency_maps}b) selects a simpler pattern with fewer, more clearly isolated modal cores. Thus, the morphology of the localized field is strongly frequency dependent within the same late-time localization band.

The frequency-resolved maps also highlight the importance of the non-overlapping granular morphology of the medium: the strongest enhancements occur in narrow inter-particle gaps but the field also penetrates the adjacent dielectric particles (Fig. \ref{frequency_maps}(c,d)). Therefore, the modal cores are not merely gap hot spots. This indicates resonant participation of small particle neighborhoods in the collective quasi-mode. Particles located in the low-intensity separation regions remain only weakly excited. On the logarithmic scale, the contrast between bright cores and dark separations exceeds two orders of magnitude, showing that neighboring modal domains are only weakly coupled to each other and to the residual background field. We note that, also in the frequency-resolved maps, the dark channels do not simply follow air--dielectric interfaces, supporting their interpretation as interference-defined separations.

We emphasize that Fig.~\ref{frequency_maps} shows only two-dimensional slices through a fully three-dimensional slab. Nevertheless, both frequency windows reveal the same qualitative organization: compact bright modal domains separated by extended low-intensity channels. Together with the time-resolved near-field maps, this demonstrates that the late-time field is self-organized into an LL-like set of weakly coupled modal basins. The exact basin pattern, however, is spectrally selective. The dynamics observed in the late-time Video 1 can therefore be interpreted as temporal competition and beating between different frequency-dependent modal morphologies within the localization window ($kr\simeq0.62-0.78$).



The persistence of the low-intensity channels and their large peak-to-valley contrast suggest a real-space mechanism for transport suppression: neighboring bright modal cores are separated by interference minima that reduce modal overlap. This morphology is reminiscent of LL theory, where landscape barriers partition a disordered medium into weakly coupled allowed regions \cite{FilocheMayboroda2012,Filoche2024}. We use this analogy phenomenologically as a direct comparison would require a vector, polarization-dependent landscape formulation for electromagnetic waves in dense dielectric media.

At this stage, we can formulate a general picture of the localization dynamics. Time filtering selects long-lived quasi-modes whose coherent superposition produces persistent low-intensity interfaces. In this regime, energy continues to leak from localized modes and through the open boundaries, but long-range internal transport is suppressed; energy decays locally rather than being carried across the system.
Unlike symmetric point scatterers \cite{Skipetrov2014}, dense packings of irregular particles create a broad distribution of gaps, shapes, and resonant neighborhoods, strongly modifying both longitudinal and transverse near-field components.
The relevant condition is therefore not merely strong disorder, but a strongly scattering and geometrically complex refractive-index environment. This is consistent with the broader view that transport in dense disordered photonic media is governed by local geometrical organization, short-range correlations, and near-field coupling \cite{Schertel2019, Aubry2017, Naraghi_etal2015, Grynko2020}. The precise role of wavelength-scale local complexity should be studied in future work through controlled comparisons of different geometrical models.


In conclusion, our results indicate that strong localization of light in 3D dielectric disorder can be identified by the convergence of dynamic, spectral, and real-space signatures. At late times, after extended diffusive components have decayed, the field reveals long-lived confined modes, non-exponential transmission, spectrally isolated resonances, and persistent low-intensity channels that separate compact modal basins. These observations show that localization in a vector Maxwell system is not only a transport effect, but also has a directly visible spatial morphology. The interference-separated, landscape-like confinement pattern provides a real-space picture linking time-resolved transport parameters to the organization of weakly coupled confinement domains. More generally, the results show that dense dielectric disorder can support localized vector electromagnetic modes through the combined action of multiple scattering, near-field coupling, and wavelength-scale geometrical complexity. This motivates future experimental studies of 3D light localization in realistic dielectric systems.


\begin{acknowledgments}
  The authors gratefully acknowledge the computing time granted by the Paderborn Center for Parallel Computing (PC$^2$).
\end{acknowledgments}

\section*{DATA AVAILABILITY}
The raw numerical data underlying all figures in this article are openly available at \url{https://doi.org/10.5281/zenodo.20937859}.







\appendix
\counterwithin{figure}{section}

\section{Supplemental Material}

\subsection{Numerical method}\label{Numerical_method}
The numerical simulations were carried out with a self-developed light scattering code based on the open-source DGTD Maxwell solver MIDG published by Tim Warburton \cite{midg_code}. The code extends MIDG with material models, boundary conditions, and different light sources. Continuous and Gaussian pulses are supported. Plane-wave excitation is implemented using the total-field/scattered-field technique. The computational domain is periodic in the transverse X and Y directions, and perfectly matched layers (PML) are applied in the longitudinal Z direction. For time-resolved transmission measurements, the sample layers are illuminated by a linearly polarized short plane-wave pulse propagating along Z. The transmitted intensity $T(t)$ is recorded in a monitor plane behind the layer, and all intensity curves are normalized by the peak intensity of the transmitted pulse. 

The Thouless conductance was extracted from the late-time transmission spectra by fitting individual resonances with Lorentzian line shapes. For each spectrum, local maxima above the background level were identified and fitted in a narrow frequency interval around the peak using a least-squares routine. The full width at half maximum obtained from the Lorentzian fit was used as the resonance linewidth \(\delta \omega\). Nearest-neighbor spacings \(\Delta \omega\) were calculated from the fitted peak positions. The Thouless conductance was then evaluated for each mode as \(g_i=\delta \omega_i/\Delta \omega_i\), and the reported value \(\langle g\rangle\) corresponds to the average over all fitted resonances in the late-time spectral window. We note that the Fourier resolution of the late-time window is finite. Therefore, peaks whose fitted linewidths approached the resolution limit were excluded from the estimate. Since finite-window broadening can only increase the apparent linewidths of the remaining resonances, the reported $\langle g\rangle$ values should be regarded as conservative upper estimates of the spectral overlap.

Particulate samples with different dimensions and controlled filling fractions were generated using the \textit{Bullet} Physics engine \cite{coumans2021}, an open-source C++ library also available as a Blender add-on. It allows time-domain simulations of the dynamics and collisions of multiple arbitrary three-dimensional shapes, enabling the construction of powder-like samples with realistic topologies. Pre-generated random irregular particles were initially distributed sparsely and then allowed to settle freely under gravity onto a substrate inside a closed rectangular volume, forming slabs with controlled volume fraction. The particle shapes were produced with a self-developed generator based on the Gaussian random field approach. The resulting three-dimensional layer models were then used for tetrahedral mesh generation. In the dielectric regions with n=3.0, the spatial resolution was approximately 15 tetrahedral cells per central wavelength of the incident pulse, and the DGTD nodal expansion order was N=3. 

All DGTD simulations were performed on the Noctua 2 HPC cluster at Paderborn University. A representative $X_L = 16$ simulation used approximately $4\times10^6$ tetrahedral elements and $6 \times 10^7$ DGTD degrees of freedom, and was run on 12 CPU nodes for 120 hours, corresponding to 184,320 core-hours.

\subsection{Late-time modal dynamics and persistence}\label{Supplement_video}

In this section we present animated dynamics of the near field recorded for a sample with thickness $X_L=16$ at late times ($t\approx 140t_a$) in the longitudinal, XZ, and transverse, XY, planes. Videos 1 and 2 show time evolution of the bright separated quasi-modes, each having its own eigenfrequency and spatial extent. The long-lived localized hotspot regions may temporarily fade and later reappear at the same spatial position. This behavior is consistent with beating between a small number of nearly degenerate localized quasi-modes. The confinement region itself remains pinned, while the local intensity is modulated by the evolving relative phases of the contributing modes populating it. Therefore, the underlying localized state is not destroyed, but becomes temporarily suppressed in intensity due to destructive interference within a weakly overlapping environment of long-lived modes.

The late-time videos also provide a qualitative check against significant reflections from the PML. The recorded XZ maps cover the full computational domain, including the regions between the slab and the absorbing boundaries. In a short-pulse simulation, any appreciable PML feedback would appear as delayed incoming wave fronts or extended boundary-related structures. Instead, Video 1 shows no waves returning from the absorbing layers toward the sample; the late-time field consists of compact resonances pinned inside the disordered slab. 

\begin{video}[H]
  \centering
  \href{https://drive.google.com/file/d/1g2PrGqfZ7ejiXgy2yZMWkBAzfm3k2uaF/view?usp=share_link} {\includegraphics[width=8cm]{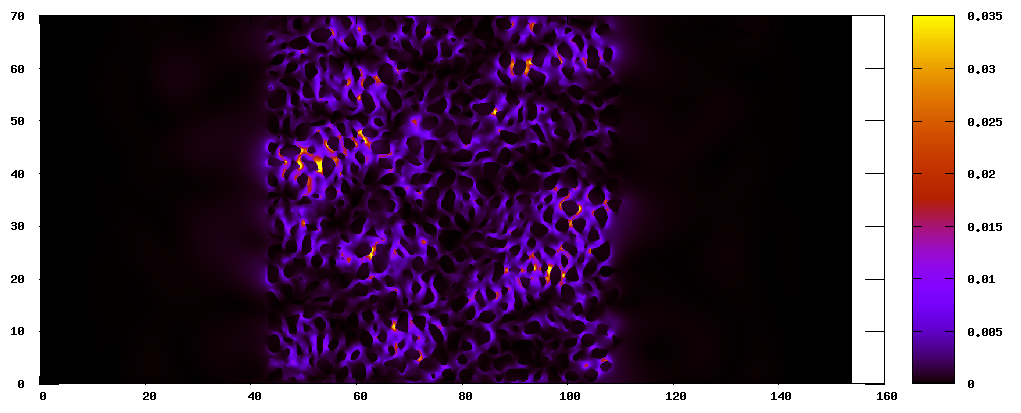}}
  \caption{  Late-time near-field dynamics in XZ plane. }
 \label{video_xz}
\end{video}

\begin{video}[H]
  \centering
  \href{https://drive.google.com/file/d/1f6bp1puevzCLKvvq1HoQEcqyif3Vm_3K/view?usp=share_link}{\includegraphics[width=5cm]{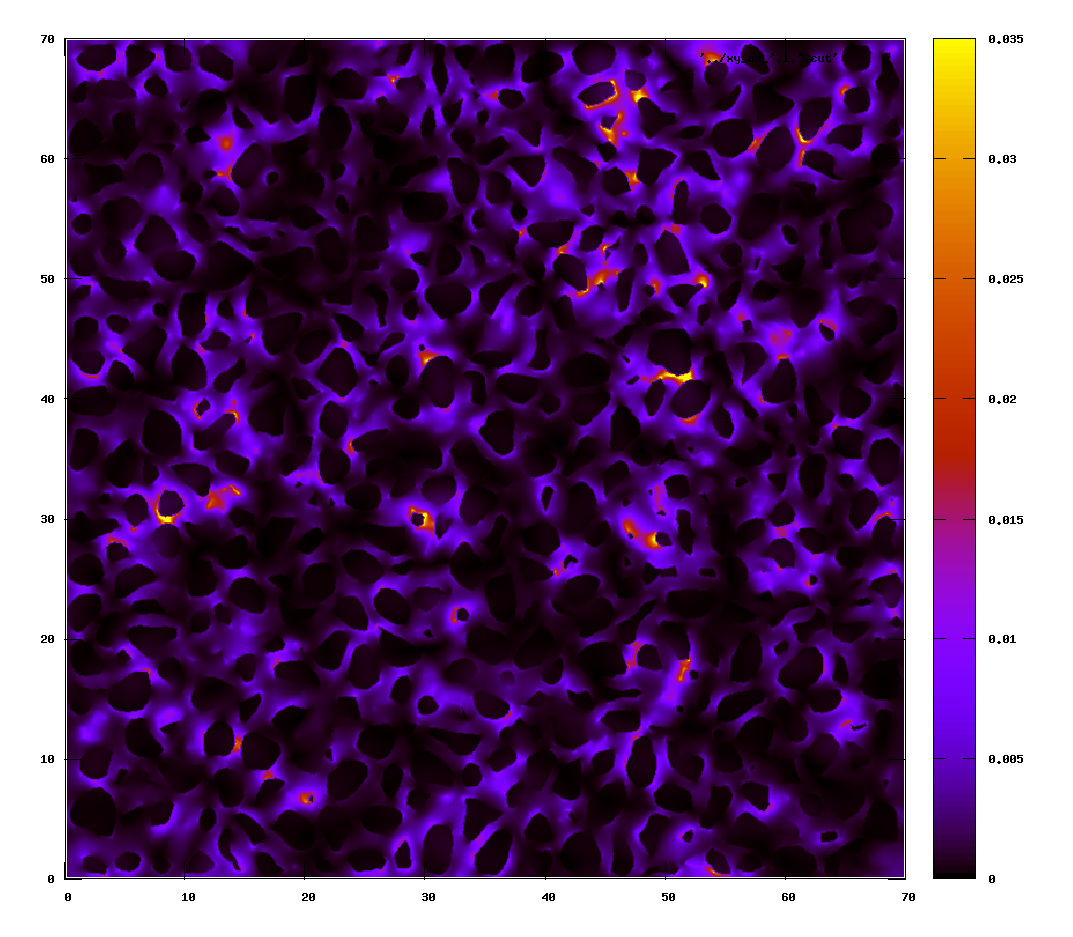}}
  \caption{  Late-time near-field dynamics in XY plane. }
  \label{video_xy}
\end{video} 

\begin{figure}[htbp]
  \centering
  \includegraphics[width=8cm]{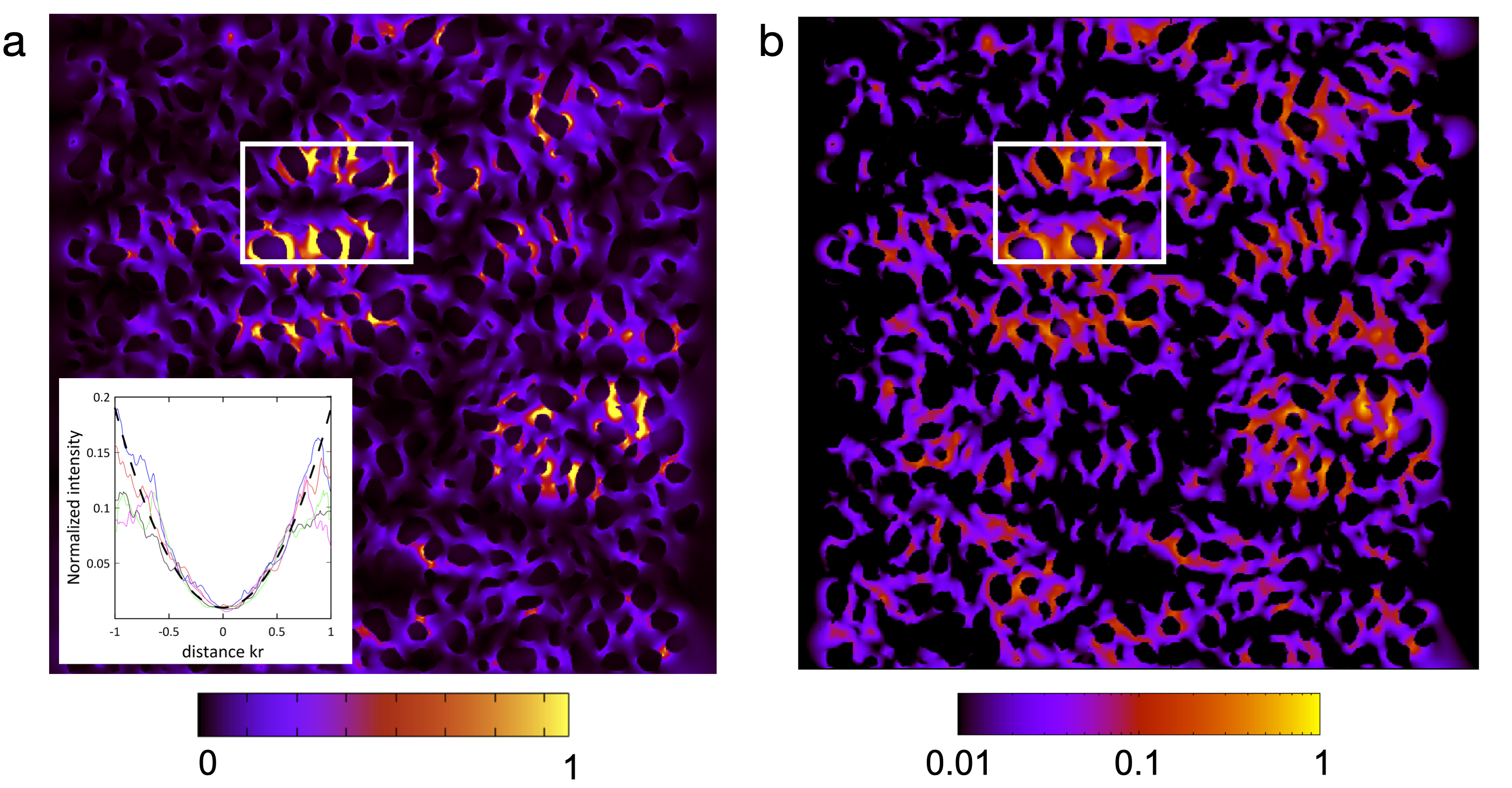}
  \caption{Interference-defined barriers between late-time modal domains.
  a, b, Late-time near-field intensity maps showing a close pair of bright localized regions separated by a dark valley, plotted on linear and logarithmic scales. a, inset, Transverse profiles across five independently selected dark valleys, each averaged over neighboring cuts. The profiles have similar widths and minima levels and are approximately parabolic near the minima. 
 }
  \label{valley_profiles}
\end{figure}

At late times, the near field breaks into spatially localized bright islands separated by persistent dark valleys. Measurements across several independently selected valleys, each averaged over 30–40 neighboring cuts, reveal a U-shaped barrier profile with similar width and minima levels at different locations (Fig. \ref{valley_profiles}a), demonstrating that these minima are not accidental fluctuations. Even after averaging, the barrier contrast remains a factor of  10-20 (Fig. \ref{valley_profiles}a, inset), while the local contrast between barrier minima and neighboring mode maxima exceeds two orders of magnitude (Fig. \ref{valley_profiles}b). 
The transverse dark-valley profiles are approximately parabolic (Fig. \ref{valley_profiles}a, inset), which is consistent with a local quadratic expansion around smooth destructive-interference minima.

\begin{figure}
\centering
\includegraphics[width=1\linewidth]{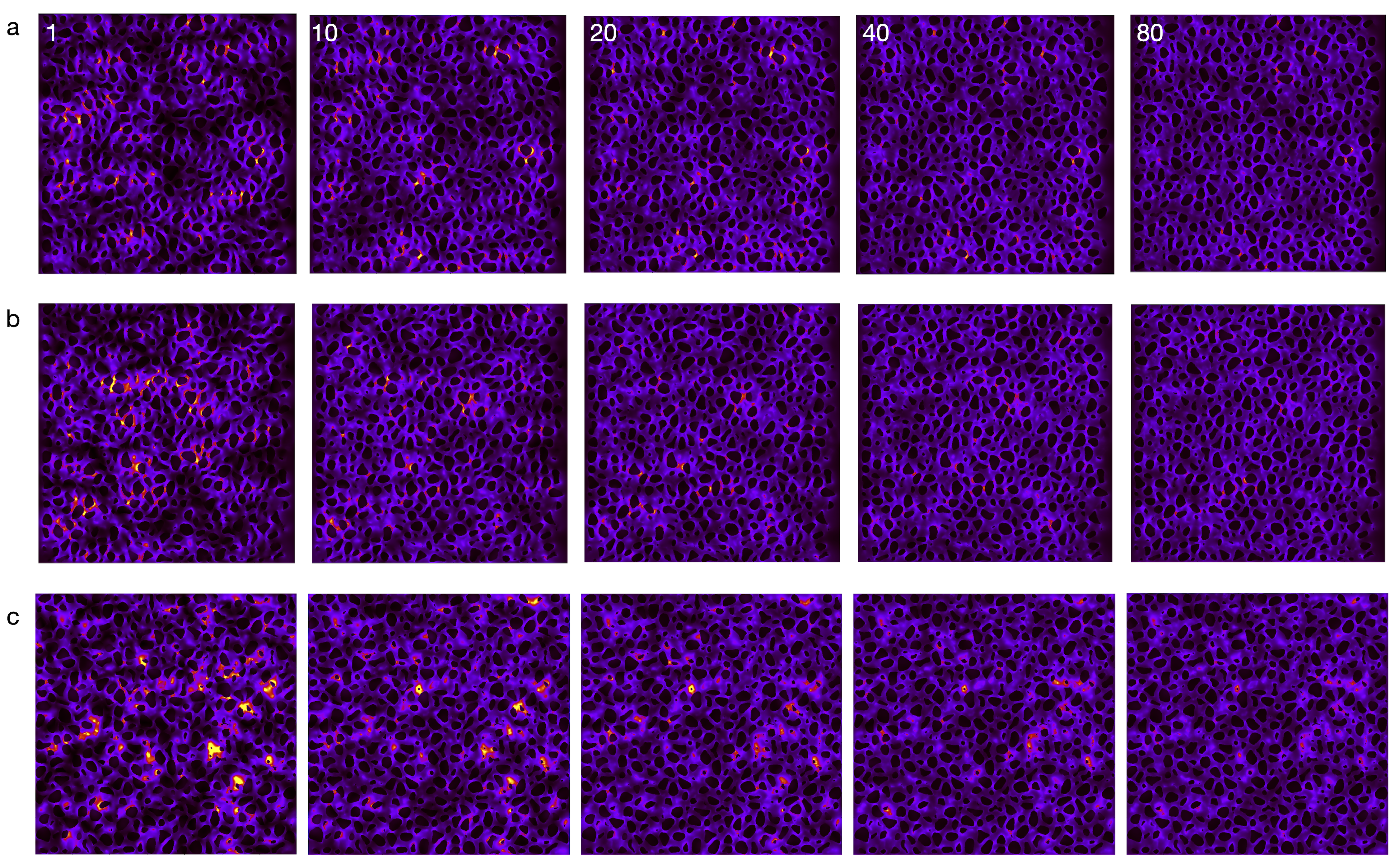}
\caption{Three-dimensional persistence of the confinement pattern. Late-time near-field intensity maps in the XZ (a), YZ (b) and XY (c) planes, averaged over 1--80 optical cycles at $t=140t_a$.
}
\label{average_fields_XZ_YZ_XY}
\end{figure}

To quantify the persistence of the late-time field structure, we compute cycle-averaged field maps and their correlations. Fig. \ref{average_fields_XZ_YZ_XY} shows intensity maps for different domain cross-sections averaged over 1–80 cycles starting at $t\approx 140t_a$. Increasing the averaging window suppresses phase-sensitive fluctuations while preserving the low-intensity regions.

\begin{figure}[htbp]
\centering
\includegraphics[width=4.5cm]{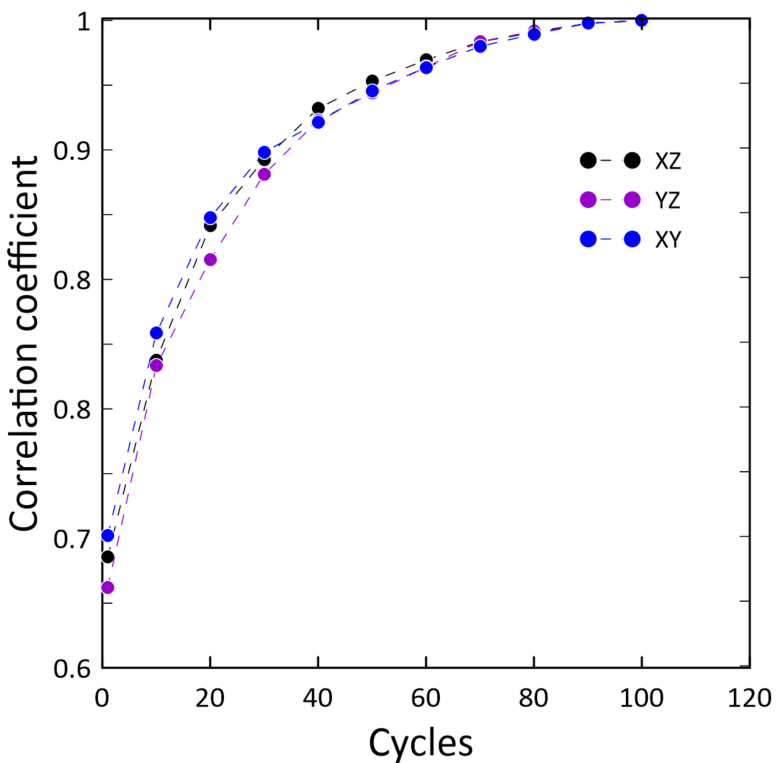}
\caption{Correlation coefficient between cycle-averaged near-field maps and the corresponding 100-cycle average, calculated for XZ, YZ and XY cross-sections.
}
\label{map_correlation}
\end{figure}

Fig. \ref{map_correlation} shows Pearson correlation coefficient calculated for different average maps with respect to the 100-cycle average. It increases monotonically with the averaging window in all three cross-section planes. For single-cycle maps the correlation is only about 0.65–0.70, indicating substantial instantaneous oscillatory or multimodal fluctuations. After averaging over 20–40 cycles, the correlation already exceeds approximately 0.85–0.93, and it approaches unity for 80–100 cycles. This shows that the late-time field has a stable underlying spatial morphology in three dimensions within the given time window.


\begin{figure}[htbp]
  \centering
  \includegraphics[width=8cm]{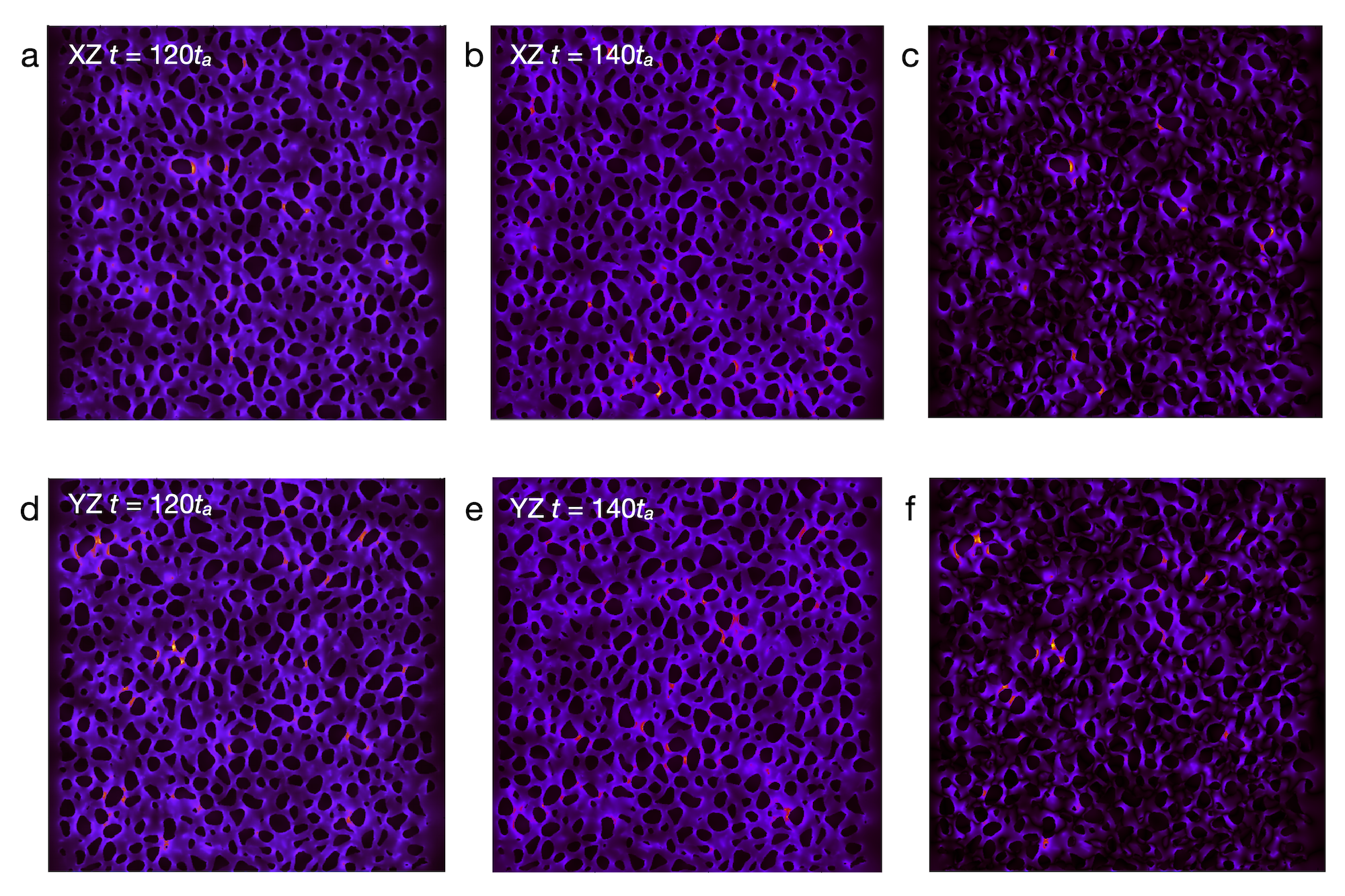}
  \caption{Long-delay persistence of the dark-channel pattern. a,b, XZ near-field intensity maps averaged over 45-cycle periods at $t\simeq120t_a$ and $t\simeq140t_a$, respectively. d,e, Corresponding YZ maps for the same two time windows. c,f, Unsigned difference maps for the XZ and YZ planes. 
  }
  \label{averages_different_times}
\end{figure}

To assess the long-term stability of the pattern, we compute analogous cycle-averaged maps at an earlier time, $t\approx 120t_a$. Fig. \ref{averages_different_times} shows comparison of 45-cycle averages separated by about 360 optical cycles. Although the maps do not show direct visual similarity, the dark regions in the difference maps (Fig. \ref{averages_different_times}(c,f)) point to areas that remain dark in both corresponding cycle-averaged field distributions. The correlation coefficient between the maps at $t\approx 120t_a$ and $t\approx 140t_a$ is $\approx 0.79$ for both the XZ and YZ planes, quantifying a persistent, though reduced, similarity of the confinement pattern.

We note that the fixed particle arrangement does contribute a nonzero baseline correlation, as confirmed by the comparison between an instantaneous snapshot and a 40-cycle average of the same realization, which typically yields correlation coefficients of only $\approx 0.3$. This provides an estimate of the purely static structural contribution. (Uncorrelated geometries like XZ and YZ cross-sections give correlation coefficients of about 0.1-0.2). The substantially higher correlation obtained between two 45-cycle averages separated by a longer period, therefore, cannot be explained by the particle geometry alone and indicates the persistence of an additional field-organized pattern. That is, late-time field organization is not regenerated randomly, but retains a substantial sample-specific confinement pattern over long delays and does not imply a complete reorganization of the confinement structure, while the observed evolution is dominated by local redistribution of modal population.


\subsection{Thickness dependence of late-time localization signatures}\label{Thickness_test}

\begin{figure}[htbp]
  \centering
  \includegraphics[width=8cm]{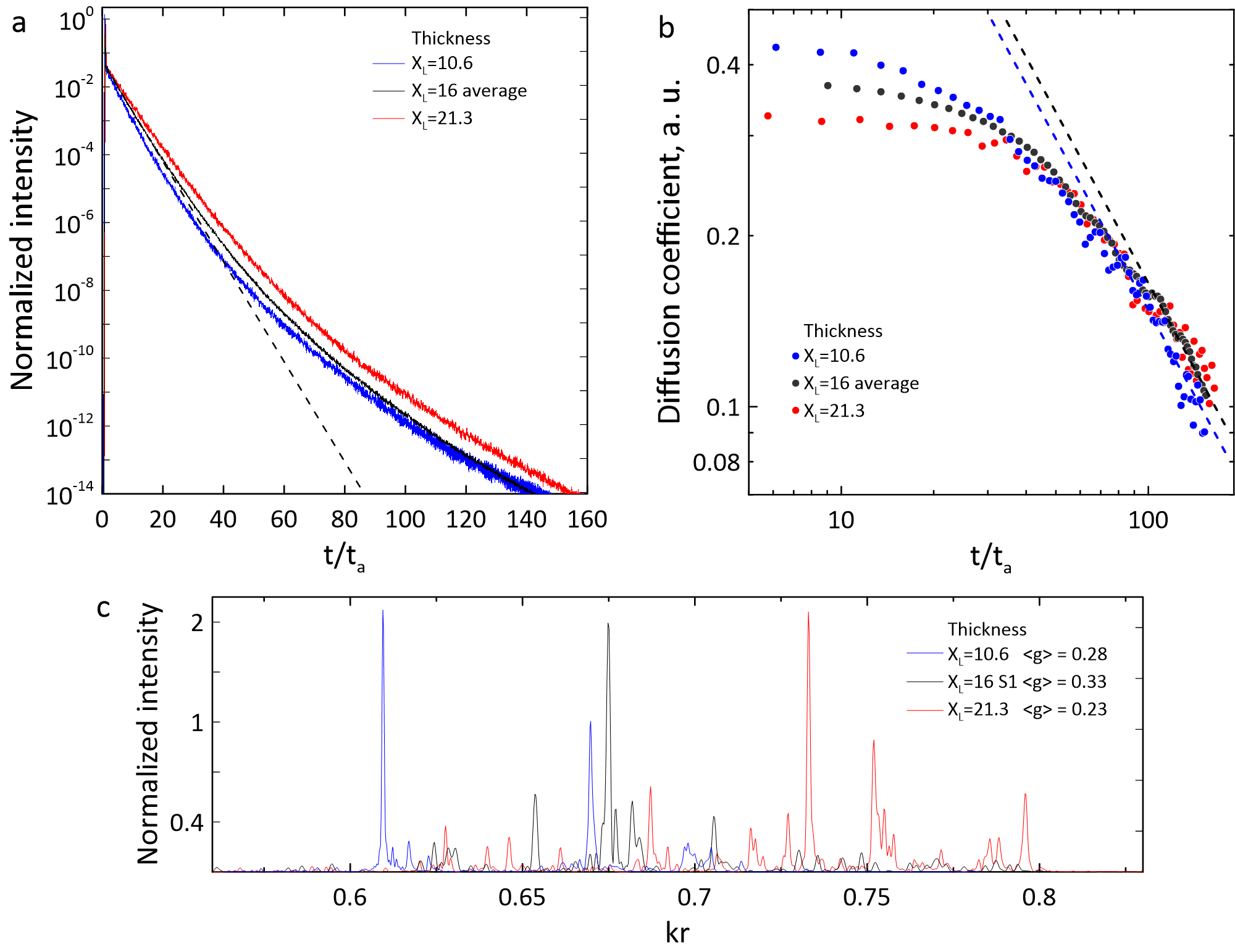}
  \caption{Thickness dependence of late-time transport and spectral signatures.
  a, Time-resolved transmission $T(t)$ for slabs with thicknesses $X_L = 10.6$, $16$ (averaged), $21.3$,  with the transverse dimensions kept fixed.  All cases develop a non-exponential late-time tail, although the temporal onset depends on thickness because extended and diffusive components escape at different rates.
  b, Time-dependent diffusion coefficient $D(t)$. 
  c,  Late-time transmission spectra and the corresponding Thouless conductances $\langle g\rangle$ calculated for the three thicknesses. 
  }
  \label{thickness_test_TDSpectra}
\end{figure}

In this section, we test the effect of slab thickness by simulating layers thinner and thicker than those considered in the main text. The transverse dimensions are kept fixed, while the thicknesses are $X_L = 10.6$ and $X_L = 21.3$ with particle numbers $N=6000$ and $N=12000$, respectively. Fig. \ref{thickness_test_TDSpectra}a compares the corresponding transmissions $T(t)$ with the average curve from Fig. \ref{Tt_Dt_Spectra}a. In all cases, a non-exponential tail emerges at long times, although the early-time $D(t)$ differ as expected for slabs of different thickness. Reducing slab thickness accelerates dynamical separation between rapidly escaping extended or diffusive components and long-lived spatially confined resonances. Once the slab remains several confinement lengths thick, localized cores can still be accommodated, while the open boundaries more efficiently suppress competing extended contributions. In thicker slabs, the larger volume supports more slowly decaying diffuse modes, so the same late-time confined regime emerges only after longer time filtering. Thus, changing $X_L$ mainly changes the time required for the extended background to leak out. Correspondingly, $D(t)$ shows the early-time variation with slab thickness, but all cases undergo the same qualitative late-time slowdown following the same downward trend toward the localized-regime scaling (Fig. \ref{thickness_test_TDSpectra}b).

The late-time spectra for all thicknesses consist of narrow resonances concentrated in the same spectral window $kr \approx 0.62 - 0.8$ and the spectral structure remains qualitatively the same (Fig. \ref{thickness_test_TDSpectra}c). The extracted Thouless conductances, $\langle g\rangle = 0.28$ ($X_L = 10.6$) and $\langle g\rangle = 0.23$ ($X_L = 21.3$) are all below unity, indicating weak modal overlap for each case. This supports the interpretation that the long-lived resonances belong to the same localized spectral band, selected by the disordered vector Maxwell operator, rather than being determined primarily by the slab thickness. The small non-monotonic variation of $\langle g\rangle$ is attributed to finite-sample modal statistics rather than a systematic thickness trend.

\begin{figure}[htbp]
  \centering
  \includegraphics[width=8.5cm]{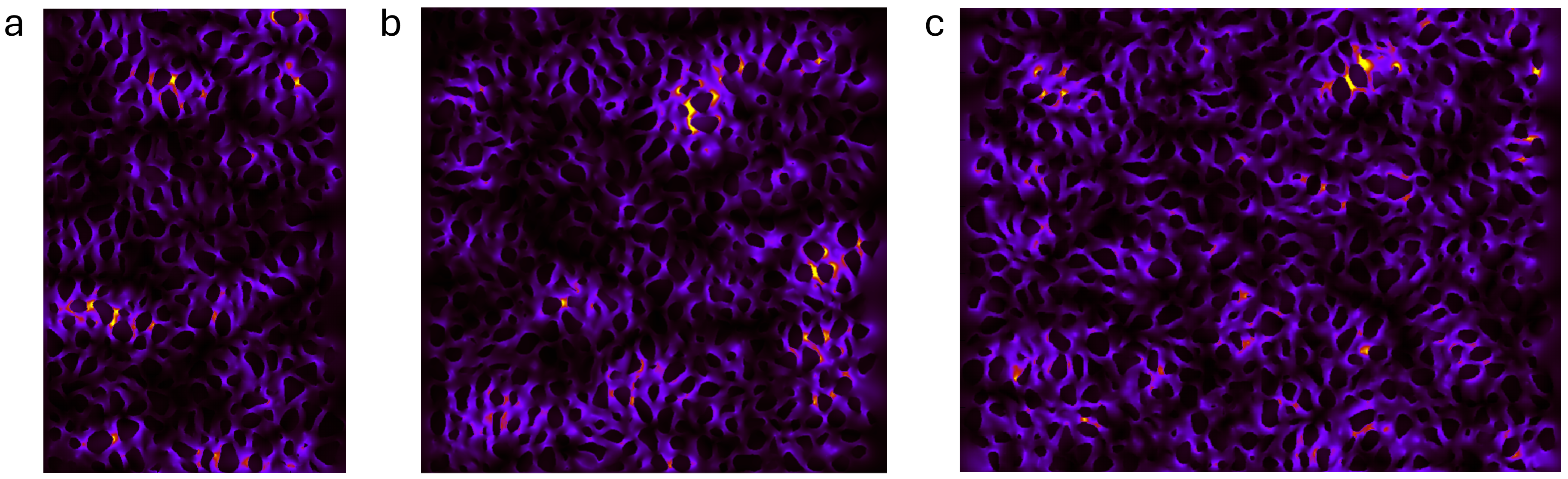}
  \caption{ Late-time near-field XZ maps showing modal organization in slabs with different thicknesses, $X_L = 10.6$ (a), $X_L=16$ (b) and $X_L = 21.3$ (c). The maps are taken at different times chosen such that $T(t)\sim10^{-13}$ for each sample.
  }
  \label{thickness_test_fields}
\end{figure}

The near-field maps in Fig. \ref{thickness_test_fields} show similar late-time confinement morphology for $X_L = 10.6$, $X_L=16$ and $X_L = 21.3$. This indicates that the resonant structures are not set by the slab thickness, but by a local confinement scale in the dense medium. The finite slab therefore acts as an open-system temporal filter: rapidly escaping components disappear first, revealing long-lived resonances whose spatial extent remains nearly thickness independent. This test does not replace a full asymptotic scaling analysis, but it supports the interpretation that the observed modal isolation is governed primarily by the disordered bulk rather than by the slab thickness alone. We note that a dedicated sample-size study would further refine the results by determining the asymptotic scaling behavior and locating the mobility edge. This would require ensemble averaging over many disorder realizations and several substantially larger system sizes and is therefore beyond the scope of the present solution. Here, we focus on the experimentally relevant finite open system and show that the late-time regime exhibits a converging set of localization signatures.

In addition, Fig. \ref{thickness_test_fields} provides evidence that localization-like confinement morphology and interference-barrier networks can emerge on mesoscopic length scales of only a few confinement lengths. This does not require a macroscopically thick random medium or very long diffusive paths. Instead, it can be initiated locally by highly resonant complex scattering neighborhoods and then stabilized by surrounding destructive-interference valleys. This points to the importance of wavelength-scale geometrical complexity in the formation of the confinement landscape. Here, complexity means that the medium contains many distinct local scattering environments: narrow gaps, irregular particle shapes, locally resonant clusters, tortuous high-index/air interfaces, and a nontrivial network of possible field paths. Such local structure may promote stronger modal confinement than simpler random models based on spherical particles or point scatterers, even at comparable nominal disorder strength.

\bibliography{bibliography}

@article{Anderson1958,
  title = {Absence of Diffusion in Certain Random Lattices},
  author = {Anderson, P. W.},
  journal = {Phys. Rev.},
  volume = {109},
  issue = {5},
  pages = {1492--1505},
  numpages = {0},
  year = {1958},
  month = {Mar},
  publisher = {American Physical Society},
  doi = {10.1103/PhysRev.109.1492},
  url = {https://link.aps.org/doi/10.1103/PhysRev.109.1492}
}

@article{grynko2023,
      title={3D Anderson localization of light in disordered systems of dielectric particles}, 
      author={Yevgen Grynko and Dustin Siebert and Jan Sperling and Jens Förstner},
      year={2023},
      eprint={2312.14393},
      archivePrefix={arXiv},
      journal = {https://arxiv.org/abs/2312.14393},
      primaryClass={physics.optics},
      url={https://arxiv.org/abs/2312.14393}, 
}

@article{THOULESS197493,
title = {Electrons in disordered systems and the theory of localization},
journal = {Physics Reports},
volume = {13},
number = {3},
pages = {93-142},
year = {1974},
issn = {0370-1573},
doi = {https://doi.org/10.1016/0370-1573(74)90029-5},
url = {https://www.sciencedirect.com/science/article/pii/0370157374900295},
author = {D.J. Thouless}
}

@article{Mondal2019,
  title = {Optical Thouless conductance and level-spacing statistics in two-dimensional Anderson localizing systems},
  author = {Mondal, Sandip and Kumar, Randhir and Kamp, Martin and Mujumdar, Sushil},
  journal = {Phys. Rev. B},
  volume = {100},
  issue = {6},
  pages = {060201},
  numpages = {5},
  year = {2019},
  month = {Aug},
  publisher = {American Physical Society},
  doi = {10.1103/PhysRevB.100.060201},
  url = {https://link.aps.org/doi/10.1103/PhysRevB.100.060201}
}

@article{EscalanteSkipetrov2018,
	author = {Escalante, Jose M. and Skipetrov, Sergey E.},
	date = {2018/08/01},
	doi = {10.1038/s41598-018-29996-1},
	id = {Escalante2018},
	isbn = {2045-2322},
	journal = {Scientific Reports},
	number = {1},
	pages = {11569},
	title = {Level spacing statistics for light in two-dimensional disordered photonic crystals},
	url = {https://doi.org/10.1038/s41598-018-29996-1},
	volume = {8},
	year = {2018}}

@article{Wiersma1997,
  author = {Wiersma, Diederik S. and Bartolini, Paolo and Lagendijk, Ad and Righini, Roberto},
  date = {1997/12/01},
  doi = {10.1038/37757},
  id = {Wiersma1997},
  isbn = {1476-4687},
  journal = {Nature},
  number = {6661},
  pages = {671--673},
  title = {Localization of light in a disordered medium},
  url = {https://doi.org/10.1038/37757},
  volume = {390},
  year = {1997}}

@article{Stoerzer2006,
  title = {Observation of the Critical Regime Near Anderson Localization of Light},
  author = {St\"orzer, Martin and Gross, Peter and Aegerter, Christof M. and Maret, Georg},
  journal = {Phys. Rev. Lett.},
  volume = {96},
  issue = {6},
  pages = {063904},
  numpages = {4},
  year = {2006},
  month = {Feb},
  publisher = {American Physical Society},
  doi = {10.1103/PhysRevLett.96.063904},
  url = {https://link.aps.org/doi/10.1103/PhysRevLett.96.063904}
}

@article{Sperling2013,
  author = {Sperling, T. and B{\"u}hrer, W. and Aegerter, C. M. and Maret, G.},
  date = {2013/01/01},
  doi = {10.1038/nphoton.2012.313},
  id = {Sperling2013},
  isbn = {1749-4893},
  journal = {Nature Photonics},
  number = {1},
  pages = {48--52},
  title = {Direct determination of the transition to localization of light in three dimensions},
  url = {https://doi.org/10.1038/nphoton.2012.313},
  volume = {7},
  year = {2013}}

@article{LAGENDIJK1996,
title = {Resonant multiple scattering of light},
journal = {Physics Reports},
volume = {270},
number = {3},
pages = {143-215},
year = {1996},
issn = {0370-1573},
doi = {https://doi.org/10.1016/0370-1573(95)00065-8},
url = {https://www.sciencedirect.com/science/article/pii/0370157395000658},
author = {Ad Lagendijk and Bart A. {van Tiggelen}}
}

@article{Scheffold1999,
  author = {Scheffold, Frank and Lenke, Ralf and Tweer, Ralf and Maret, Georg},
  date = {1999/03/01},
  doi = {10.1038/18347},
  id = {Scheffold1999},
  isbn = {1476-4687},
  journal = {Nature},
  number = {6724},
  pages = {206--207},
  title = {Localization or classical diffusion of light?},
  url = {https://doi.org/10.1038/18347},
  volume = {398},
  year = {1999}}

@article{Scheffold2013,
  author = {Scheffold, Frank and Wiersma, Diederik},
  date = {2013/12/01},
  doi = {10.1038/nphoton.2013.210},
  id = {Scheffold2013},
  isbn = {1749-4893},
  journal = {Nature Photonics},
  number = {12},
  pages = {934--934},
  title = {Inelastic scattering puts in question recent claims of Anderson localization of light},
  url = {https://doi.org/10.1038/nphoton.2013.210},
  volume = {7},
  year = {2013}}

@article{Sperling_2016,
  author = {T Sperling and L Schertel and M Ackermann and G J Aubry and C M Aegerter and G Maret},
  doi = {10.1088/1367-2630/18/1/013039},
  journal = {New Journal of Physics},
  month = {jan},
  number = {1},
  pages = {013039},
  publisher = {IOP Publishing},
  title = {Can 3D light localization be reached in `white paint'?},
  url = {https://dx.doi.org/10.1088/1367-2630/18/1/013039},
  volume = {18},
  year = {2016}}

@article{Skipetrov_2016,
  author = {S E Skipetrov and J H Page},
  doi = {10.1088/1367-2630/18/2/021001},
  journal = {New Journal of Physics},
  month = {jan},
  number = {2},
  pages = {021001},
  publisher = {IOP Publishing},
  title = {Red light for Anderson localization},
  url = {https://dx.doi.org/10.1088/1367-2630/18/2/021001},
  volume = {18},
  year = {2016}}

@article{SkipetrovVanTiggelen2006,
  title = {Dynamics of Anderson Localization in Open 3D Media},
  author = {Skipetrov, S. E. and van Tiggelen, B. A.},
  journal = {Phys. Rev. Lett.},
  volume = {96},
  issue = {4},
  pages = {043902},
  numpages = {4},
  year = {2006},
  month = {Feb},
  publisher = {American Physical Society},
  doi = {10.1103/PhysRevLett.96.043902},
  url = {https://link.aps.org/doi/10.1103/PhysRevLett.96.043902}
}

@article{Grynko2020,
  author = {Yevgen Grynko and Yuriy Shkuratov and Jens F{\"o}rstner},
  doi = {https://doi.org/10.1016/j.jqsrt.2020.107234},
  issn = {0022-4073},
  journal = {Journal of Quantitative Spectroscopy and Radiative Transfer},
  pages = {107234},
  title = {Light backscattering from large clusters of densely packed irregular particles},
  url = {https://www.sciencedirect.com/science/article/pii/S0022407320305148},
  volume = {255},
  year = {2020}
}

@inbook{IoffeRegel1960,
address = {New York},
  author = {A. F. Ioffe and A. R. Regel},
  booktitle = {Progress in Semiconductors},
  editor = {A. F. Gibson},
  pages = {237--291},
  publisher = {Wiley},
  title = {Non-crystalline, amorphous, and liquid electronic semiconductors},
volume = {4},
  year = {1960}}

@inbook{Grynko_chapter_2022,
  address = {Cham},
  author = {Grynko, Yevgen and Shkuratov, Yuriy and Alhaddad, Samer and F{\"o}rstner, Jens},
  booktitle = {Springer Series in Light Scattering: Volume 8: Light Polarization and Multiple Scattering in Turbid Media},
  doi = {10.1007/978-3-031-10298-1_4},
  editor = {Kokhanovsky, Alexander},
  isbn = {978-3-031-10298-1},
  pages = {125--155},
  publisher = {Springer International Publishing},
  title = {Light Scattering by Large Densely Packed Clusters of Particles},
  url = {https://doi.org/10.1007/978-3-031-10298-1_4},
  year = {2022}}

@article{Grynko2018,
	author = {Yevgen Grynko and Yuriy Shkuratov and Jens F\"{o}rstner},
	journal = {Opt. Lett.},
	month = {Aug},
	number = {15},
	pages = {3562--3565},
	title = {Intensity surge and negative polarization of light from compact irregular particles},
	volume = {43},
	year = {2018}}

@MISC{coumans2021,
author =   {Erwin Coumans and Yunfei Bai},
title =    {PyBullet, a Python module for physics simulation for games, robotics and machine learning},
howpublished = {\url{http://pybullet.org}},
year = {2016--2021}
}

@MISC{midg_code,
author =   {Tim Warburton},
title =    {Mini Discontinuous Galerkin Maxwells Time-domain Solver},
howpublished = {\url{https://github.com/tcew}},
}

@article{Skipetrov2014,
  title = {Absence of Anderson Localization of Light in a Random Ensemble of Point Scatterers},
  author = {Skipetrov, S. E. and Sokolov, I. M.},
  journal = {Phys. Rev. Lett.},
  volume = {112},
  issue = {2},
  pages = {023905},
  numpages = {5},
  year = {2014},
  month = {Jan},
  publisher = {American Physical Society},
  doi = {10.1103/PhysRevLett.112.023905},
  url = {https://link.aps.org/doi/10.1103/PhysRevLett.112.023905}
}

@article{SkipetrovSokolov2018,
  title = {Ioffe-Regel criterion for Anderson localization in the model of resonant point scatterers},
  author = {Skipetrov, S. E. and Sokolov, I. M.},
  journal = {Phys. Rev. B},
  volume = {98},
  issue = {6},
  pages = {064207},
  numpages = {8},
  year = {2018},
  month = {Aug},
  publisher = {American Physical Society},
  doi = {10.1103/PhysRevB.98.064207},
  url = {https://link.aps.org/doi/10.1103/PhysRevB.98.064207}
}

@article{Naraghi_etal2015,
  title = {Near-Field Effects in Mesoscopic Light Transport},
  author = {Rezvani Naraghi, R. and Sukhov, S. and S\'aenz, J. J. and Dogariu, A.},
  journal = {Phys. Rev. Lett.},
  volume = {115},
  issue = {20},
  pages = {203903},
  numpages = {5},
  year = {2015},
  month = {Nov},
  publisher = {American Physical Society},
  doi = {10.1103/PhysRevLett.115.203903},
  url = {https://link.aps.org/doi/10.1103/PhysRevLett.115.203903}
}

@article{Yamilov_etal2022,
  author = {Yamilov, Alexey and Skipetrov, Sergey E. and Hughes, Tyler W. and Minkov, Momchil and Yu, Zongfu and Cao, Hui},
  journal = {Nature Physics},
  number = {9},
  pages = {1308--1313},
  title = {Anderson localization of electromagnetic waves in three dimensions},
  volume = {19},
  year = {2023}}

@article{Haberko2020,
  author = {Haberko, Jakub and Froufe-P{\'e}rez, Luis S. and Scheffold, Frank},
  journal = {Nature Communications},
  number = {1},
  pages = {4867},
  title = {Transition from light diffusion to localization in three-dimensional amorphous dielectric networks near the band edge},
  volume = {11},
  year = {2020}}

@article{HESTHAVEN2002,
  author = {J.S Hesthaven and T Warburton},
  doi = {https://doi.org/10.1006/jcph.2002.7118},
  issn = {0021-9991},
  journal = {Journal of Computational Physics},
  number = {1},
  pages = {186-221},
  title = {Nodal High-Order Methods on Unstructured Grids: I. Time-Domain Solution of Maxwell's Equations},
  url = {https://www.sciencedirect.com/science/article/pii/S0021999102971184},
  volume = {181},
  year = {2002}}

@article{Hefei2008,
	author = {Hu, Hefei and Strybulevych, A. and Page, J. H. and Skipetrov, S. E. and van Tiggelen, B. A.},
	journal = {Nature Physics},
	number = {12},
	pages = {945--948},
	title = {Localization of ultrasound in a three-dimensional elastic network},
	volume = {4},
	year = {2008}}

@article{Kondov2011,
	author = {S. S. Kondov and W. R. McGehee and J. J. Zirbel and B. DeMarco},
	journal = {Science},
	number = {6052},
	pages = {66-68},
	title = {Three-Dimensional Anderson Localization of Ultracold Matter},
	volume = {334},
	year = {2011}}

@article{Crespi2013,
	author = {Crespi, Andrea and Osellame, Roberto and Ramponi, Roberta and Giovannetti, Vittorio and Fazio, Rosario and Sansoni, Linda and De Nicola, Francesco and Sciarrino, Fabio and Mataloni, Paolo},
	journal = {Nature Photonics},
	number = {4},
	pages = {322--328},
	title = {Anderson localization of entangled photons in an integrated quantum walk},
	volume = {7},
	year = {2013}}

@article{Schwartz2007,
	author = {Schwartz, Tal and Bartal, Guy and Fishman, Shmuel and Segev, Mordechai},
	journal = {Nature},
	number = {7131},
	pages = {52--55},
	title = {Transport and Anderson localization in disordered two-dimensional photonic lattices},
	volume = {446},
	year = {2007}}

@article{Riboli2011,
author = {F. Riboli and P. Barthelemy and S. Vignolini and F. Intonti and A. De Rossi and S. Combrie and D. S. Wiersma},
journal = {Opt. Lett.},
number = {2},
pages = {127--129},
publisher = {Optica Publishing Group},
title = {Anderson localization of near-visible light in two dimensions},
volume = {36},
month = {Jan},
year = {2011},
url = {https://opg.optica.org/ol/abstract.cfm?URI=ol-36-2-127},
doi = {10.1364/OL.36.000127},
}

@article{FilocheMayboroda2012,
author = {Marcel Filoche  and Svitlana Mayboroda },
title = {Universal mechanism for Anderson and weak localization},
journal = {Proceedings of the National Academy of Sciences},
volume = {109},
number = {37},
pages = {14761-14766},
year = {2012},
doi = {10.1073/pnas.1120432109},
URL = {https://www.pnas.org/doi/abs/10.1073/pnas.1120432109},
eprint = {https://www.pnas.org/doi/pdf/10.1073/pnas.1120432109}
}

@article{Skipetrov2024,
  title = {Higher-order localization landscape theory of Anderson localization},
  author = {Skipetrov, Sergey E.},
  journal = {Phys. Rev. B},
  volume = {110},
  issue = {21},
  pages = {214209},
  numpages = {10},
  year = {2024},
  month = {Dec},
  publisher = {American Physical Society},
  doi = {10.1103/PhysRevB.110.214209},
  url = {https://link.aps.org/doi/10.1103/PhysRevB.110.214209}
}

@article{Filoche2024,
  title = {Anderson mobility edge as a percolation transition},
  author = {Filoche, Marcel and Pelletier, Pierre and Delande, Dominique and Mayboroda, Svitlana},
  journal = {Phys. Rev. B},
  volume = {109},
  issue = {22},
  pages = {L220202},
  numpages = {6},
  year = {2024},
  month = {Jun},
  publisher = {American Physical Society},
  doi = {10.1103/PhysRevB.109.L220202},
  url = {https://link.aps.org/doi/10.1103/PhysRevB.109.L220202}
}

@article{ Lyra2015,
  author = {{Lyra, M. L.} and {Mayboroda, S.} and {Filoche, M.}},
  title = {Dual landscapes in Anderson localization on discrete lattices},
  DOI= "10.1209/0295-5075/109/47001",
  url= "https://doi.org/10.1209/0295-5075/109/47001",
  journal = {EPL},
  year = 2015,
  volume = 109,
  number = 4,
  pages = "47001",
  month = "",
}

@article{Shubitidze24,
author = {Tornike Shubitidze and Yilin Zhu and Hari Sundar and Luca Dal Negro},
journal = {Opt. Mater. Express},
number = {4},
pages = {1008--1024},
publisher = {Optica Publishing Group},
title = {Localization landscape of optical waves in multifractal photonic membranes},
volume = {14},
month = {Apr},
year = {2024},
url = {https://opg.optica.org/ome/abstract.cfm?URI=ome-14-4-1008},
doi = {10.1364/OME.520201},
}

@article{Rashidi21,
author = {Mohammad Rashidi and Ziyuan Li and Chennupati Jagadish and Sudha Mokkapati and Hark Hoe Tan},
journal = {Opt. Express},
number = {21},
pages = {33548--33557},
publisher = {Optica Publishing Group},
title = {Controlling the lasing modes in random lasers operating in the Anderson localization regime},
volume = {29},
month = {Oct},
year = {2021},
url = {https://opg.optica.org/oe/abstract.cfm?URI=oe-29-21-33548},
doi = {10.1364/OE.441003},
}

@article{Vasco_Hughes2018,
  author = {Vasco, Juan Pablo and Hughes, Stephen},
  date = {2018/04/18},
  doi = {10.1021/acsphotonics.7b00967},
  journal = {ACS Photonics},
  journal1 = {ACS Photonics},
  journal2 = {ACS Photonics},
  month = {04},
  number = {4},
  pages = {1262--1272},
  publisher = {American Chemical Society},
  title = {Anderson Localization in Disordered LN Photonic Crystal Slab Cavities},
  type = {doi: 10.1021/acsphotonics.7b00967},
  url = {https://doi.org/10.1021/acsphotonics.7b00967},
  volume = {5},
  year = {2018},
  year1 = {2018}}

@article{Trojak2017,
    author = {Trojak, Oliver Joe and Crane, Tom and Sapienza, Luca},
    title = {Optical sensing with Anderson-localised light},
    journal = {Applied Physics Letters},
    volume = {111},
    number = {14},
    pages = {141103},
    year = {2017},
    month = {10},
    issn = {0003-6951},
    doi = {10.1063/1.4999936},
    url = {https://doi.org/10.1063/1.4999936}
}

@article{Arregui2023,
  title = {Cavity Optomechanics with Anderson-Localized Optical Modes},
  author = {Arregui, G. and Ng, R. C. and Albrechtsen, M. and Stobbe, S. and Sotomayor-Torres, C. M. and Garc\'{\i}a, P. D.},
  journal = {Phys. Rev. Lett.},
  volume = {130},
  issue = {4},
  pages = {043802},
  numpages = {7},
  year = {2023},
  month = {Jan},
  publisher = {American Physical Society},
  doi = {10.1103/PhysRevLett.130.043802},
  url = {https://link.aps.org/doi/10.1103/PhysRevLett.130.043802}
}

@article{Laurent2007,
  title = {Localized Modes in a Finite-Size Open Disordered Microwave Cavity},
  author = {Laurent, David and Legrand, Olivier and Sebbah, Patrick and Vanneste, Christian and Mortessagne, Fabrice},
  journal = {Phys. Rev. Lett.},
  volume = {99},
  issue = {25},
  pages = {253902},
  numpages = {4},
  year = {2007},
  month = {Dec},
  publisher = {American Physical Society},
  doi = {10.1103/PhysRevLett.99.253902},
  url = {https://link.aps.org/doi/10.1103/PhysRevLett.99.253902}
}

@phdthesis{RazoLopez2024Localization,
  author       = {Razo L{\'o}pez, Luis Alberto},
  title        = {Localization of Electromagnetic Waves Beyond Anderson: Role of Correlations, Symmetries and Topology},
  school       = {Universit{\'e} C{\^o}te d'Azur},
  year         = {2024},
  type         = {PhD thesis},
  address      = {Nice, France},
  url          = {https://theses.hal.science/tel-04633536v1},
  hal_id       = {tel-04633536},
  note         = {NNT: 2024COAZ5013}
}

@article{Molinari2012,
author = {Diego Molinari and Andrea Fratalocchi},
journal = {Opt. Express},
number = {16},
pages = {18156--18164},
publisher = {Optica Publishing Group},
title = {Route to strong localization of light: the role of disorder},
volume = {20},
month = {Jul},
year = {2012},
url = {https://opg.optica.org/oe/abstract.cfm?URI=oe-20-16-18156},
doi = {10.1364/OE.20.018156},
}

@article{Xianling2014,
  title = {Anderson localization at the subwavelength scale for surface plasmon polaritons in disordered arrays of metallic nanowires},
  author = {Shi, Xianling and Chen, Xianfeng and Malomed, Boris A. and Panoiu, Nicolae C. and Ye, Fangwei},
  journal = {Phys. Rev. B},
  volume = {89},
  issue = {19},
  pages = {195428},
  numpages = {5},
  year = {2014},
  month = {May},
  publisher = {American Physical Society},
  doi = {10.1103/PhysRevB.89.195428},
  url = {https://link.aps.org/doi/10.1103/PhysRevB.89.195428}
}

@article{Aubry2020,
  title = {Experimental Tuning of Transport Regimes in Hyperuniform Disordered Photonic Materials},
  author = {Aubry, Geoffroy J. and Froufe-P\'erez, Luis S. and Kuhl, Ulrich and Legrand, Olivier and Scheffold, Frank and Mortessagne, Fabrice},
  journal = {Phys. Rev. Lett.},
  volume = {125},
  issue = {12},
  pages = {127402},
  numpages = {6},
  year = {2020},
  month = {Sep},
  publisher = {American Physical Society},
  doi = {10.1103/PhysRevLett.125.127402},
  url = {https://link.aps.org/doi/10.1103/PhysRevLett.125.127402}
}

@article{Dalichaouch1991,
  author = {Dalichaouch, Rachida and Armstrong, J. P. and Schultz, S. and Platzman, P. M. and McCall, S. L.},
  date = {1991/11/01},
  doi = {10.1038/354053a0},
  id = {Dalichaouch1991},
  isbn = {1476-4687},
  journal = {Nature},
  number = {6348},
  pages = {53--55},
  title = {Microwave localization by two-dimensional random scattering},
  url = {https://doi.org/10.1038/354053a0},
  volume = {354},
  year = {1991}}

@article{Aubry2017,
  title = {Resonant transport and near-field effects in photonic glasses},
  author = {Aubry, Geoffroy J. and Schertel, Lukas and Chen, Mengdi and Weyer, Henrik and Aegerter, Christof M. and Polarz, Sebastian and C\"olfen, Helmut and Maret, Georg},
  journal = {Phys. Rev. A},
  volume = {96},
  issue = {4},
  pages = {043871},
  numpages = {7},
  year = {2017},
  month = {Oct},
  publisher = {American Physical Society},
  doi = {10.1103/PhysRevA.96.043871},
  url = {https://link.aps.org/doi/10.1103/PhysRevA.96.043871}
}

@article{Schertel2019,
author = {Schertel, Lukas and Siedentop, Lukas and Meijer, Janne-Mieke and Keim, Peter and Aegerter, Christof M. and Aubry, Geoffroy J. and Maret, Georg},
title = {The Structural Colors of Photonic Glasses},
journal = {Advanced Optical Materials},
volume = {7},
number = {15},
pages = {1900442},
doi = {https://doi.org/10.1002/adom.201900442},
url = {https://advanced.onlinelibrary.wiley.com/doi/abs/10.1002/adom.201900442},
eprint = {https://advanced.onlinelibrary.wiley.com/doi/pdf/10.1002/adom.201900442},
year = {2019}
}

@article{Conley2014,
  title = {Light Transport and Localization in Two-Dimensional Correlated Disorder},
  author = {Conley, Gaurasundar M. and Burresi, Matteo and Pratesi, Filippo and Vynck, Kevin and Wiersma, Diederik S.},
  journal = {Phys. Rev. Lett.},
  volume = {112},
  issue = {14},
  pages = {143901},
  numpages = {5},
  year = {2014},
  month = {Apr},
  publisher = {American Physical Society},
  doi = {10.1103/PhysRevLett.112.143901},
  url = {https://link.aps.org/doi/10.1103/PhysRevLett.112.143901}
}

@article{Rafayelyan2020,
  author = {Rafayelyan, Mushegh and Dong, Jonathan and Tan, Yongqi and Krzakala, Florent and Gigan, Sylvain},
  title = {Large-Scale Optical Reservoir Computing for Spatiotemporal Chaotic Systems Prediction},
  journal = {Physical Review X},
  volume = {10},
  pages = {041037},
  year = {2020},
  doi = {10.1103/PhysRevX.10.041037}
}

@article{Wang2024,
  author = {Wang, Hao and Hu, Jianqi and Morandi, Andrea and Nardi, Alfonso and Xia, Fei and Li, Xuanchen and Savo, Romolo and Liu, Qiang and Grange, Rachel and Gigan, Sylvain},
  title = {Large-scale photonic computing with nonlinear disordered media},
  journal = {Nature Computational Science},
  year = {2024},
  volume = {4},
  number = {6},
  pages = {429--439},
  doi = {10.1038/s43588-024-00644-1},
  url = {https://doi.org/10.1038/s43588-024-00644-1},
  issn = {2662-8457}
}

@article{Caselli2017,
  author  = {Caselli, Niccolò and Intonti, Francesca and La China, Federico and Biccari, Francesco and Riboli, Francesco and Gerardino, Annamaria and Li, Lianhe and Linfield, Edmund H. and Pagliano, Francesco and Fiore, Andrea and Gurioli, Massimo},
  title   = {Near-field speckle imaging of light localization in disordered photonic systems},
  journal = {Applied Physics Letters},
  year    = {2017},
  volume  = {110},
  number  = {8},
  pages   = {081102},
  doi     = {10.1063/1.4976747},
  url     = {https://doi.org/10.1063/1.4976747},
  issn    = {0003-6951}
}

@article{Abrahams1979,
  title = {Scaling Theory of Localization: Absence of Quantum Diffusion in Two Dimensions},
  author = {Abrahams, E. and Anderson, P. W. and Licciardello, D. C. and Ramakrishnan, T. V.},
  journal = {Physical Review Letters},
  volume = {42},
  pages = {673--676},
  year = {1979},
  doi = {10.1103/PhysRevLett.42.673}
}

\end{document}